# From the Wobble to Reliable Hypothesis
Semenov D.A. (dasem@mail.ru)

**Abstract.** A simple explanation for the symmetry and degeneracy of the genetic code has been suggested. An alternative to the wobble hypothesis has been proposed. This hypothesis offers explanations for: i) the difference between thymine and uracil, ii) encoding of tryptophan by only one codon, iii) why E. coli have no inosine in isoleucine tRNA, but isoleucine is encoded by three codons. The facts revealed in this study offer a new insight into physical mechanisms of the functioning of the genetic code.

The wobble hypothesis, which was first proposed more than 40 years ago [1], can explain two events: 1) formation of the uracil-guanine pair as a result of codon-anticodon interaction in the third position and 2) encoding of isoleucine by three codons (for the universal genetic code). This hypothesis is based on two statements: in the third position of the codon nucleotides can "wobble", and inosine can form pairs with uracil, cytosine, and adenine. Neither of these statements is necessary for explaining the abovementioned events, and the wobble hypothesis is just wrong. Once we admit this, we can get a better insight into the structure and functioning of the genetic code.

Crick's hypothesis completely ignores two chemical factors. The first is the stacking between two neighboring nucleotides, and the second is the ability of uracil to be present in the enol form.

What makes the uracil-guanine pair possible? Uracil can exist either in the enol form or in the keto form. Uracil in the enol form can make two and even three hydrogen bonds with guanine, while uracil in the keto form would be able to make only one hydrogen bond. One can say that the enol form of uracil is stabilized by its interaction with the complementary guanine.

The molecular basis of the wobble hypothesis has been criticized by other researchers [2, 3], but I'd like to add a few comments.

Watson first considered nucleotide formulae in the enol form as the more probable, guided by reference books of that time [4]. Fortunately, I can give convincing arguments using commonly available information.

Codon-anticodon interaction results in the formation of a short segment of the double helix. In the case of some codons, opposite to uracil in the third position of the codon there is inosine in the anticodon. Crick's wobble hypothesis [1] allows a solution for this pair only by wobbling the third nucleotides of the codon and the anticodon. However, in the double helix, their position is also stabilized by stacking. If stacking exerts significant influence so that wobbling becomes impossible, then no uracil-inosine complementarity is possible for the keto form of uracil, i.e. there cannot be any hydrogen bonds. If in this case uracil is in the enol form, in the state that most closely imitates cytosine, two hydrogen bonds can be established, without a nucleotide shift. We do not have to invent a new kind of complementary pairs in order to comprehend the formation of the G-U pair. If uracil is in the enol form, Watson-Crick pairs should suffice.

If C-I are Watson-Crick's pairs and U-I pairs are formed in accordance with the wobble hypothesis, it is not clear why these codons are always indistinguishable, although they have different conformations, which can be stabilized. Crick's hypothesis would be valid if cytosine and uracil were opposed by the same anticodon but as a component of different tRNAs, which would stabilize different conformations of the anticodon. This, however, is not so – the same tRNA always "confuses" cytosine with uracil in the third position of the codon. With the keto-enol tautomerism, both codons have the same conformation.

Why is uracil opposed by inosine in some anticodons and by guanine in others? According to Crick's hypothesis, inosine and guanine are indistinguishable to uracil. If we suppose the presence of the enol form of uracil, we can suggest that inosine emerged in anticodons in an evolutionary way because in this case the guanine amino group could not form a hydrogen bond. That is, in this case inosine (guanine without an amino group) is sufficient. The weakening and possible disruption of the hydrogen bond that has been formed involving the amino group must cause more than just binding of vacant hydrogen bonds to water molecules. Binding of two water molecules will transform the amino group from an active helper into an active obstacle.

I should add that the enol form of uracil can be registered in NMR spectra, which makes the necessary experiments easy to perform.

Incorporation of thymine into DNA may be accounted for as follows. Methylation could favor further stabilization of uracil in the keto form. Oxygen is an electron acceptor while the methyl group is an electron donor, so the presence of the methyl group reduces the probability of the proton being near oxygen [5, 6]. The wobble hypothesis does not discriminate between uracil and thymine. Neither do contemporary textbooks for school and university students. A direct experiment seems to be an obvious way to test this: mRNA with the poly-(UCG) structure can encode poly-Ser, poly-Val, poly-Arg, depending on the position of the reading-frame. My hypothesis suggests that synthesis of poly-Arg cannot occur on the mRNA matrix with the poly-(TCG) structure, as thymine cannot construct the enol form.

Let us consider deamination of cytosine in DNA. It is a source of G-U pairs in DNA. G-U pair in DNA it is pair with enol form of uracil. About 50% of cytosine in DNA is methylated and deamination converts cytosine to thymine. G-T pairs are effectively repaired because thymine does not construct an enol form.

The emergence of thymine prevented the ambiguity caused by the chemical properties of uracil.

Confirmation of my hypothesis can be found in studies systematically addressing uracil derivatives in the wobble position.

The influence of the derivative in the 5 position of uracil on the construction of the enol form was addressed in a number of studies: Sowers and colleagues [7] suggested that the mutagenic activity of 5-bromouracil is related to the relative ease of the formation of the enol form.

Thymine and uracil are considered to be nucleotides with similar properties, while 5-bromouracil is assumed to be different from either of them. The fact that the difference between thymine and uracil in their ability to construct the enol form is consistently ignored prevents one from realizing that DNA contains thymine and RNA uracil. My interpretation of this is as follows: thymine is present in the keto form only and can pair up but with adenine. Uracil, on the other hand, can easily construct the enol form; that is why it forms the U-G pair and for the same reason it would lead to mutagenesis, like 5-bromouracil, if it were contained in DNA.

Zeegers-Huyskens [5] theoretically demonstrated the influence of derivatives in the 5 position, confirming that the presence of the electron-accepting derivative decreases the basicity of oxygen in the 4 position. Of special interest is the result for thymine: the donor derivative (methyl group) qualitatively changes the result – the basicity of oxygen increases dramatically. That is, it is more difficult for thymine to construct the enol form. This result places uracil in one group with its 5 substituted derivatives, if the derivative is an acceptor of σ-electron density, while thymine drops out of this group.

The difference between thymine and uracil is 3 pH units, or 1000 for the corresponding constant of oxygen atom basicity. This is a well-known difference between the stabilities of DNA and RNA in the form of a double helix.

Realization of this fact makes it possible to interpret two results immediately related to the wobble hypothesis:

Takai and his co-authors [6] demonstrated that 5-oximethyluridine in the wobble position facilitates the formation of the wobble pair.

According to Kurata and colleagues [8], the same effect was produced by 5-carboxymethylaminomethyluridine and 5-taurinomethyluridine, which contradicted the expectations of the experimenters.

Moreover, Takai and Yokoyama [9] demonstrated the absence of the proton in the 3 position of uracil under similar conditions, which is a direct proof of my hypothesis.

And finally, Näsvall and colleagues [10] state that 5-oxiacetyluridine facilitates the formation of the "wobble" pair. They proposed the existence of enol 5-oxiacetyluridine to explain this result.

Unfortunately, there don't seem to be any similarly detailed studies on uracil-inosine and uracil-guanine (with guanine in the anticodon) pairs. Moreover, studies referred to above [8, 10] suggest that only modified uracils of the anticodon can be present in the enol form, allowing the formation of the wobble pair without modification.

In each case, a new substantiation has to be invented for the results.

One can avoid this constant violation of the principle of reasoning named Occam's razor by recognizing that the U-G pair has been made by the enol form of uracil, only. Then, all the above-mentioned experimental data can be naturally interpreted as the effect of the acceptor derivative in the 5 position of uracil.

I'd like to discuss in greater detail studies addressing the inosine-adenine pair, postulated by Crick [1] to account for degeneracy of isoleucine in the table of the universal genetic code. As stated above, I consider this postulate unnecessary. As noted by Lagekvist [3], there is no need for these pairs to form: to read the codon ending in A there always is an anticodon beginning with U. Lagekvist, however, did not propose any explanation for isoleucine codons. In 2004, Murfy and Ramakrishnan reported an experimental confirmation, on the atomic-molecular level, of the existence of the inosine-adenine pair within the ribosome-tRNA-mRNA complex; moreover, the conformation of the pair was exactly the same as predicted by Crick [11]. The authors could have found what they had been looking for, but the data they presented made their search meaningless. The authors of that study experimented with arginine tRNA of *E. coli*, explaining their choice as follows: "The modification of adenine to inosine has been found for all the four-codon boxes except tRNAGly—eight tRNAs in higher eukaryotes and seven in yeast, but only tRNA Arg in prokaryotes. In *Escherichia coli* inosine is present only in tRNA Arg ICG, which is the only tRNA to decode the codons CGU/C/A". That is, *E. coli* and many other prokaryotes disprove the wobble hypothesis in direct experiment: they have no inosine in isoleucine tRNA, but isoleucine is encoded by three codons. In the second part of my study I explain this phenomenon.

Based on the above arguments, we do not have to take into account considerable nucleotide shifts. Hereafter we assume that nucleotides occupy Watson-Crick's classic positions. Thus, our attention can be focused on possible small conformational changes in the codon-anticodon pair, which do not cause a hydrogen bond rupture.

Let us now discuss the second part of Crick's hypothesis: why is isoleucine encoded by three codons? A more general question is: why are methionine and tryptophan encoded without degeneracy, by one codon each?

To answer this question, we have to refer to another study published in 1966 – Yu.B. Rumer's study of the symmetry of the genetic code table.

Let us consider the table of the genetic code letter doublets as it was proposed by Yu.B. Rumer [12, 13] (Table 1). The author focused his attention on the presence of the letter doublets (i.e. the first two nucleotides of the triplet, called "roots" by Yu.B. Rumer) and their ability or inability to encode just one amino acid. Of 16 letter doublets, 8 were strong (encoding just one amino acid) and 8 were weak (encoding more than one amino acid). Please note that this symmetry is characteristic of almost all dialects of the genetic code.

I should add that the symmetry of the genetic code is the property of the whole system of gene coding. It is an experimental fact, summarizing 64 experimental facts, one for each codon.

The keto-enol tautomerism is a good example of dramatic changes in the form and properties of certain nucleotides that occur under rather weak interactions.

A nucleotide triplet can be also assumed to change its form, but in this case a change in the form of the first two nucleotides occurs due to the third. Here I do not speak of the form of the triplet in the solution or the form of the triplet in mRNA, but rather the form of the double helix segment resulting from the codon-anticodon interaction.

This form is determined by two forces – complementary interactions and stacking (the interaction of neighboring nucleotides). Stacking is a nonspecific event as there is a well-known formula: purine-purine>purine-pyrimidine>pyrimidine-pyrimidine.

|   | C | G | U | A |
|---|---|---|---|---|
| C | Pro | Arg | Leu | His / Gln |
| G | Ala | Gly | Val | Asp / Glu |
| U | Ser | Cys / Trp/Stop | Phe / Leu | Tyr / Stop |
| A | Thr | Ser / Arg | Ile / Met | Asn / Lys |

Table 1. The symmetry of the table of the letter doublets for the genetic code (according to Rumer). Strong letter doublets are marked in gray.

In the third position of the codon the number of the formed hydrogen bonds (for the first approximation) is much less relevant than the nature of the bases – purine or pyrimidine. Thus, the presence of strong and weak letter doublets (Table 1) can naturally be related to the presence of stacking.

In codons with the CC, CG, GC, and GG letter doublets, the form of the doublet is only determined by complementary interaction. The three hydrogen bonds in each doublet make conformational changes impossible.

Let us now discuss the (UC-UG), (AC-AG), (CU-CA) and (GU-GA) letter doublets. In each doublet the first letter is strong and the second weak. The number of hydrogen bonds in each doublet is the same, but doublets with purine in the second position are the weak ones. Due to the presence of purine in the second position, two purines can occur one by one – in the second and the third positions of the codon. This construction imparts sufficient stress to the first two nucleotides to cause a change in their conformation. Please note that the anticodon is much less conformationally flexible because it is part of tRNA and is stabilized by its structure.

In codons with the UA and AA doublets, the conformation of the doublet can be changed due to the presence of purine in the third position of the codon. In the codons with the UU and AU doublets, the conformation of the codon can be changed by just a slight interaction between pyrimidine and purine.

Are there any analogs of the conformational changes described here in the available literature? Strange as it may seem, there are. RNA can form an A-form double helix or an A' double helix. They differ in the number of nucleotides per turn (11 and 12, respectively), the pitch of the helix (30 and 36 Angstroms), and slopes of nucleotides; otherwise, the conformations are similar [14]. Interestingly, in these forms, the contribution of stacking to the formation of the double helix structure differs significantly.

Essentially, I suppose that the nucleotide pairs in the codon-anticodon mini-double helix can be positioned relative to each other in such a way that the relative position of the first and second pair can be best described by the A-form (for instance) and the second and third pair – by the A'-form.

In the table of the universal genetic code there are two amino acids, each of which is encoded by one codon only: tryptophan (Trp), encoded by the UGG codon, and methionine (Met), encoded by AUG. This feature can be consistently explained based upon the same arguments that have been used to explain Rumer's symmetry. If there are letter doublets incapable of conformational changes when pyrimidine is replaced by purine in the last letter and there are letter doublets that readily change under these conditions, there must be letter doublets that are close to the equilibrium point. For these letter doublets even much smaller impacts may prove to be significant. This model can be illustrated "mechanically": one scale pan holds the total number of hydrogen bonds formed by the first two base pairs and the other – the stacking force between the second and the third nucleotides.

The UG doublet is situated on the table diagonal, i.e. it may be close to the equilibrium point. For its conformational change it may be important not only that the third position is occupied by purine but also that this is guanine. An additional hydrogen bond leads to a conformational change. The situation with the AU doublet is similar, but instead of the purine-purine interaction, we have to assume that its conformation can be changed by a weaker, pyrimidine-purine, interaction. In this case, the letter doublet itself is extremely prone to conformational changes: the doublet contains no nucleotides capable of forming three hydrogen bonds. The AU doublet is presumably close to the equilibrium point and the conformational transition is achieved at the expense of one hydrogen bond in the last nucleotide.

Thus, it is the first two nucleotides, or, to be more exact, the conformation of the first two nucleotides, that are the encoding part of the codon. For half of the letter doublets this conformation depends upon the last nucleotide, although it is this conformation rather than the whole sequence that is recognized. The number of different letter doublets of codons is not large: 8 (one for each strong doublet) + 16 (two for each weak one) = 24. There is only one codon for which all three nucleotides are significant – UGA, the stop codon. UGA's potential anticodon – UCA, placed into the tryptophan tRNA, would not be able to encode tryptophan. It would keep the conformation of the first two nucleotides unchanged. This can be a way to verify my speculations experimentally.

At the present time there are evidently no calculation procedures for molecule structure that would take into account the influence of one hydrogen bond on the conformation of the double helix with three base pairs. This may be more significant than just a temporary technical limitation. For physical reasons, biological macromolecules can be in the state sensitive to weak changes, which accounts for their functional activity. Physics is familiar with such states; this is, e.g., the state near the critical point of the second-order phase transition. Theoretically, in these states, even very weak impacts can cause significant concerted structural changes. This is attractive for technology and even more attractive for evolution, which is originally sparing in expenditure of energy resources.

How could a similar effect be expressed in the functioning of the genetic code? Let us suppose that the codon that takes the most advantageous conformation makes the anticodon accept the corresponding conformation. This effect cannot be strong as all our reasoning has been based on the limited mobility of the anticodon. The effect must be certainly weaker than the effect of the formation of one hydrogen bond; however, if it caused a change in the conformation of the whole tRNA, this would suggest its active influence on protein biosynthesis. This would also suggest a crisis in contemporary computational methods in molecular biology: what is related to biological function of macromolecules would inevitably be lost due to calculation errors.

**Conclusion.**

My hypothesis accounts for all experimental facts considered by Francis Crick in his "wobble hypothesis". Moreover, my hypothesis offers explanations for: i) the difference between thymine and uracil, ii) encoding of tryptophan by only one codon, iii) the symmetry of the genetic code, iv) the experimental results collected by Ulf Lagerkvist, and v) the difference between inosine and guanine.

**Acknowledgement.**

The author would like to thank Krasova E. for her assistance in preparing this manuscript. I wish to express my gratitude to Dr. Wolfgang Pluhar for his appropriate suggestions and encouragement.